\documentclass[fleqn,usenatbib]{mnras}

\usepackage{newtxtext,newtxmath}
\usepackage[T1]{fontenc}

\DeclareRobustCommand{\VAN}[3]{#2}
\let\VANthebibliography\thebibliography
\def\thebibliography{\DeclareRobustCommand{\VAN}[3]{##3}\VANthebibliography}

\usepackage{graphicx}	% Including figure files
\usepackage{hyperref}
\usepackage{natbib}
\usepackage{float}
\usepackage{graphicx}

\title [Strobed Imaging for Seeing Diagnostics]{Strobed Imaging as a Method for the Determination and Diagnosis of Local Seeing}

\author[C. W. Stubbs]{
Christopher W. Stubbs,$^{1}$\thanks{E-mail: stubbs@physics.harvard.edu}
\\
$^{1}$Department of Physics \& Department of Astronomy, Harvard University, Cambridge MA USA 02138\\
}

\date{Accepted XXX. Received YYY; in original form ZZZ}

\pubyear{2021}

\begin{document}
\label{firstpage}
\pagerange{\pageref{firstpage}--\pageref{lastpage}}
\maketitle

% Abstract of the paper
\begin{abstract}
The image quality budget of many telescopes can have substantial contributions from local seeing, both``mirror'' and ``dome'', which arise from turbulence and temperature variations that are difficult to quantify, measure directly, and ameliorate. We describe a method to determine the ``local'' seeing degradation due to wavefront perturbations within the final tens of meters of the optical path from celestial sources to a ground-based telescope, using the primary instrument and along the same path taken by light from celestial sources. The concept involves placing strobed emitters along the light path to produce images on the main focal plane that ``freeze'' different realizations of index perturbations.  This method has the advantage of measuring directly the image motion and scintillation imparted by the dynamic spatial and temporal structure of local perturbations in the index of refraction along the light path, with a clean separation from seeing induced in the atmosphere above the dome.  The strobed-source approach allows for rapid image motion and scintillation to be measured directly on the focal plane, even for large-aperture telescopes with wide field instruments and slow shutters, such as that being constructed for the Rubin Observatory. A conceptual design is presented that uses the ``guider'' CCDs in the Rubin telescope focal plane to make local-seeing measurements on demand, perhaps even during science exposures. 
\end{abstract}

% Select between one and six entries from the list of approved keywords.
% Don't make up new ones.

%%%%%%%%%%%%%%%%%%%%%%%%%%%%%%%%%%%%%%%%%%%%%%%%%%

%%%%%%%%%%%%%%%%% BODY OF PAPER %%%%%%%%%%%%%%%%%%

\section{Introduction}

Ground-based optical and infrared telescopes suffer from sources of image quality degradation that typically preclude achieving the diffraction limit of $\theta \sim \lambda/D$. For $D$=10 m class telescopes at optical wavelengths ($\lambda \sim $500nm), the typical achieved FWHM of around 1 arcsec for a celestial point source (absent adaptive optics correction) is about 100 times worse than the diffraction limit. 

A variety of factors contribute to this image degradation, including:
\begin{itemize}
    \item index of refraction variations in the upper atmosphere, 
    \item ``ground layer'' or ``boundary layer'' turbulence due to the boundary condition of zero wind velocity at the Earth's surface, 
    \item perturbations to laminar airflow due to local topography and structures,
    \item turbulence within the enclosure, due to ambient and driven airflow through the slit and vents in the dome, 
    \item thermally driven air currents due to power dissipation on the telescope top end, and other locations, 
    \item turbulence and thermal currents in the vicinity of the primary mirror due to temperature differences between the mirror and the adjacent air, 
    \item tracking errors and vibrations in the mirror support systems, 
    \item wind-driven oscillations and motions of the mirror support systems, and
    \item quasi-static optical aberrations in the optical system. 
\end{itemize}

The performance optimization of modern ground-based telescopes entails an iterative assessment of the image quality budget of the system and the suppression of the dominant sources of image degradation. Image quality is a major figure of merit for most observatories, and strongly determines the scientific utility of the system. The past few decades have seen considerable progress in both methodologies and understanding of this process, including active optical systems to adjust for quasi-static aberrations and adaptive optical systems to compensate for wavefront distortions due to the atmosphere. Reviews of astronomical seeing are given in  \cite{hickson2014atmospheric} and \cite{hardy1998adaptive}. Papers that pertain specifically to the management of ``local seeing,'' arising from within the telescope enclosure include \cite{woolf1979dome, zago1997engineering, racine1991mirror}. 

The design process for next-generation large-aperture ground-based telescopes (Rubin, GMT, TMT) typically includes detailed computational fluid dynamics simulations of airflow and 
turbulence within the enclosure \citep{sebag2014estimating, vogiatzis2018dome, conan2019modeling,vogiatzis2018precision}. The designers use this information as they strive to minimize local seeing effects with appropriate systems engineering design choices. 
Once these systems are built we need to learn how to operate them, including airflow and thermal management within the enclosure, so as to maximize their scientific impact. 

\section{Phenomenology}

Seeing produces both ray deflection and scintillation. The wavefront propagating through the system from a point source at infinity suffers from perturbations in both angle-of-arrival and in surface brightness. The exposure-time-averaged intensity distribution on the focal plane determines the point spread function (PSF). The ratio of angular deflection to scintillation in an arriving wavefront is an indicator of where the wavefront degradation is occurring along the line of sight to the source. This ratio is used by MASS-DIMM systems to estimate the vertical profile of the structure function of the index of refraction, $C_n^2(z)$, as described in \cite{kornilov2007combined}. 

The small-angle scattering and refraction that produce scintillation in a plane wave require a considerable propagation distance to accumulate enough multi-path interference to generate substantial fluctuations in surface brightness. If a source is moved a distance $R$ into a scintillating medium, the intensity variance in the arriving flux increases as R$^{5/6}$. Moreover, the characteristic transverse spatial scale of scintillation is the geometric mean of the wavelength $\lambda$ and the distance R to the index perturbation. This is the Fresnel length, $F=\sqrt{\lambda R}$. For optical wavelengths and nearby scattering screens, F(R=20m)=3mm and F(R=1m)=700 $\mu$m. Attempting to measure scintillation over apertures larger than $F$ will dilute the signature, due to ``aperture averaging.'' Another potential source of concern for near-field scintillation is whether the range-dependent Fresnel length scale is smaller than the inner scale of the turbulent eddies. The phase shifts induced by optical path length differences are wavelength-dependent, so scintillation is best probed using short wavelengths, with an optically-narrowband source. Measuring the transverse spatial scale of scintillation to infer the distance to the perturbation is the principle behind the FASS seeing diagnostic system \citep{guesalaga2021fass}. The temporal coherence timescale is another diagnostic observable, but is not typically exploited. 

Beam deflection, by comparison, is not as suppressed by short-distance effects. Any wavefront tilt introduced in a plane wave produces the same image motion on the focal plane, regardless of where (upstream of the pupil) it arises. This suggests that image motion might be more effective for diagnosing local seeing than attempting to measure scintillation. Also, measuring wavefront tilt is a more direct path to determining the point spread function (PSF) on the focal plane, as opposed to inferring $C_n^2(x,y,z)$ from sparsely sampled data, obtained outside the beam of interest, and attempting to use that to compute beam deflection and image motion statistics.  
The focal plane centroid displacement due to local seeing has two contributions, one from before the pupil and one from after the pupil where the beam is converging.  An illustration of the complex nature of ``dome seeing'' comes from comparing the effect of a screen of index of refraction perturbations placed above the top end of the telescope with one that lies between the primary and secondary mirrors. If the seeing layer is above the top end, each light ray passes through it once. The Rubin telescope is a three-mirror configuration, and if the seeing layer is instead placed between the mirrors then each light ray passes through it not once but four times before striking the focal plane. 
    
The focusing optical system of the telescope converts angle of arrival at the pupil into focal plane position. Absent any perturbations beyond the pupil or optical aberrations, the RMS fluctuations in angle-of-arrival at the pupil convert directly into the RMS width of the PSF, with the conversion factor being the focal length of the system. 

After passing through the telescope pupil the rays start to converge, and from the seeing perspective two things happen. The beam width shrinks and so for a fixed transverse gradient in refractive index the rays experience smaller differences in deflection. Second, the lever arm to convert from angular deflection to centroid displacement is reduced as the converging beam approaches the focal plane. An index variation just above the focal plane doesn't introduce much centroid motion at all. These effects combine to produce \citep{wheelon2001electromagnetic} a weighting of $z^3$ for how index perturbations induce centroid motion in a converging beam, where $z$ is the optical path distance upstream from the focal plane. 
    
We can now place this in a more quantitative framework. If the index perturbations above and below the pupil are uncorrelated, the two resulting deflection variances add. For a subpupil of size $D$, a focal length $FL$, a path-integrated distance $Z_p$ from the focal plane to the pupil, a collimated source placed at a distance $Z_s$ from the focal plane, and a spatially-uniform $C_n^2$ index perturbation structure within the enclosure,  the variance in image position is given by \citep{wheelon2001electromagnetic} 

$$<p^2>=C_n^2  D^{-1/3} [1.092  Z_p^3   + 2.914 (Z_s-Z_p) FL^2 ].~~~ \rm{(1)} $$
    
The $z$ coordinate in equation (1) runs from the focal plane outward along the multi-reflected optical path traversed by a ray, and is not the vertical elevation in the dome. The first term in square brackets applies to the converging beam after the pupil, with a z$^{\rm{3}}$ path-dependence that takes into account both the smaller turbulence scales spanned near focus and the shorter level arm for angular deflections to produce image motion. The second term in square bracket arises from the column of air between the pupil and the output of the transmitting collimator, and depends only on the integrated path traversed by the collimated beam. The image motion does not depend on where the collimated light is deflected. 
    
The temporal power spectrum of these centroid motions is dictated by the characteristic timescale $\tau \sim D_{Fried}/v$ where $D_{Fried}$ is the isoplanatic length scale and $v$ is the transverse wind speed. 

\section{Measurement Options}

Once a telescope goes into operation, a variety of methods have been used to assess the local seeing contributions within the telescope enclosure. This work to date has predominantly involved indirect methods, such as:  1) measuring the physical driving terms, through in-dome turbulence monitors or microthermometry of the ambient air, 2) correlating the delivered image FWHM with physical parameters such as mirror-to-ambient temperature differences, while attempting to distinguish local seeing from upper atmospheric effects, and 3) looking out through the slit with diagnostic instruments such as a differential image motion monitor (DIMM), and comparing the observed image motion and scintillation statistics to those obtained from an identical system operated outside the dome. \cite{lai2019direct} and \cite{bustos2018dome} describe direct optical determinations of turbulence within a dome, but those measurements are not carried out along the same light path that is traversed by celestial sources and can miss, for example, the contribution from mirror seeing.  \cite{bridgeland1997measurements} describe time-resolved thermometry above the mirror of the William Herschel Telescope to assess mirror seeing, but converting from temperature gradients at a few points to a prediction of image degradation is difficult.

While much has been learned from the techniques described above, they are indirect methods. The ideal situation would be to monitor the local image quality perturbations along the same light path traversed by the celestial sources of interest, since this measures exactly what we care about. Using the primary astronomical instrument to assess local seeing is difficult for a number of reasons. Using celestial sources requires that we distinguish between upper atmosphere and local contributions to seeing. 
% Point-like (uncollimated) artificial sources placed within the dome produce images with an angular size of (D/L) where D is the telescope diameter and L is the distance to the source, and basically produce a flat surface brightness on the focal plane. 
Another factor that limits the use of the main instrument is incompatibility of characteristic timescales. DIMMs operate with millisecond exposure times, since in the frozen-screen approximation the characteristic timescale is $\tau \sim D_{Fried}/V_{wind}$, where $D_{Fried}$ is the length scale over which induced phase errors are coherent. The resulting coherence timescale is of order (0.1 m/ (10 m/s)) $<$  tens of milliseconds. Exposures longer than this average over the image motion we wish to characterize. The shutters on large instruments don't typically move fast enough to take images this short, even if illuminated from a collimated light source within the dome. 

This paper presents a conceptual solution to these limitations, with a scheme that could allow the use of the primary instrument to diagnose local seeing contributions.

\section{Conceptual Designs}

\begin{figure}
    \centering
    \includegraphics[width=3.8in]{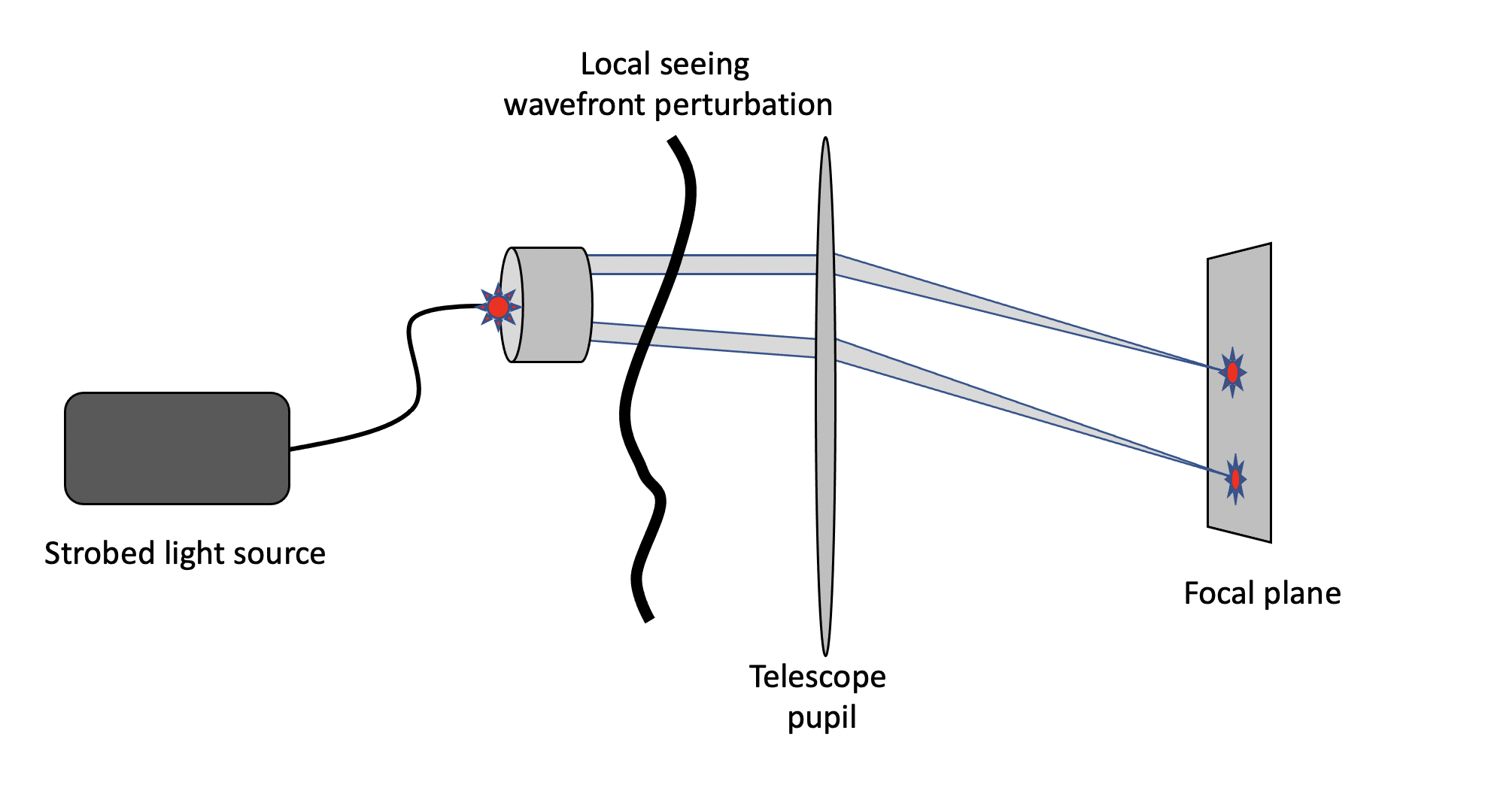}
    \caption{Strobed Local Seeing Monitor, Reverse-DIMM version. A pulsed light source illuminates the focus of a multi-aperture differential image motion monitor (DIMM) telescope, run in reverse. Placing wedges on the exit apertures steers beams 1 and 2 into different angles of arrival, separating the corresponding images on the focal plane. Measuring differential image motion probes the difference in perturbation-induced wavefront tilt due to local seeing, while being insensitive to boresight jitter or drift in either the projector or the main telescope. The short light pulses avoid issues with shutter speed for a large imager, such as that of the Rubin telescope.}
    \label{fig:rdimm}
\end{figure}

\begin{figure}
    \centering
    \includegraphics[width=3.8in]{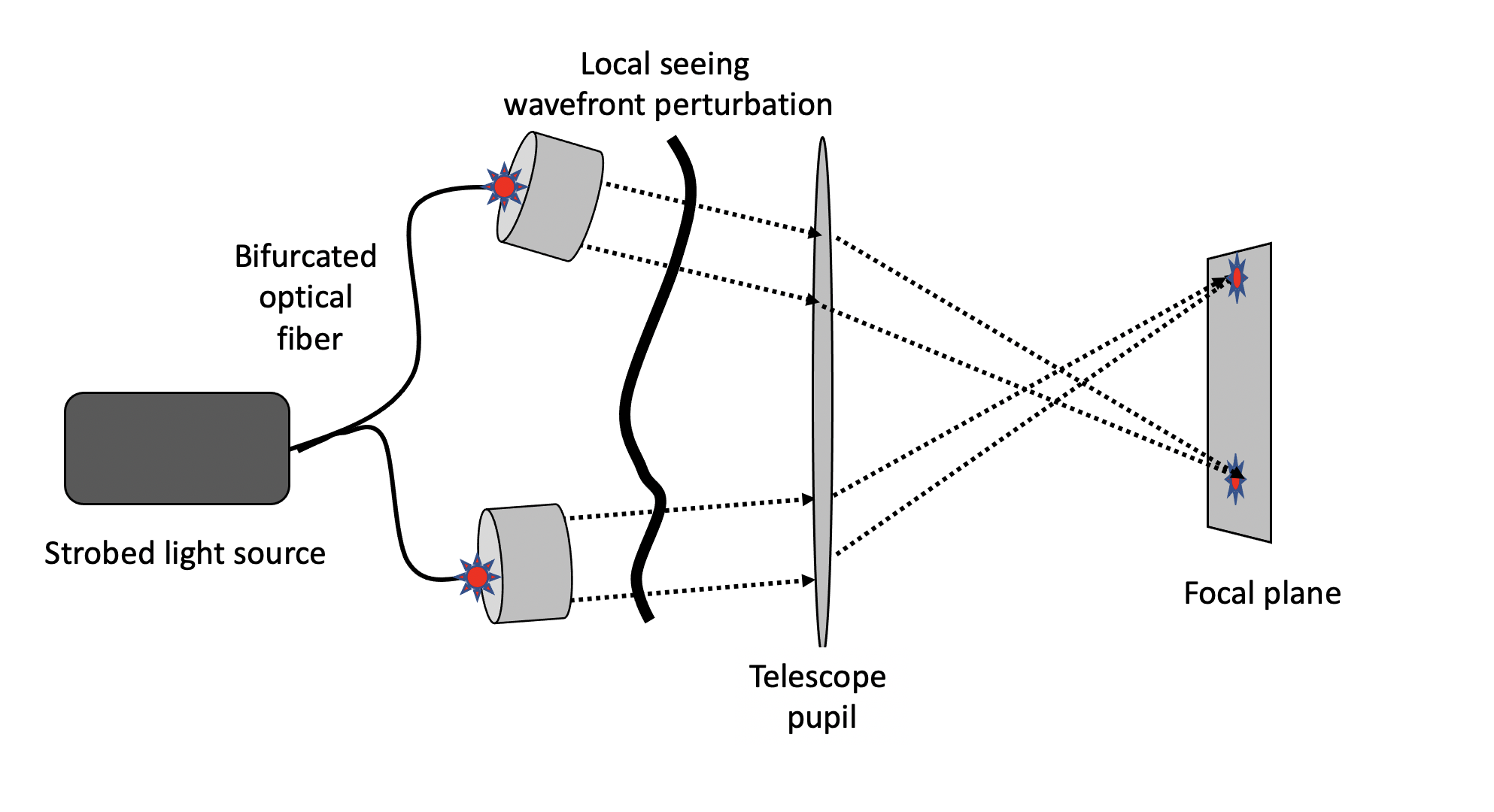}
    \caption{Strobed Local Seeing Monitor, Multi-Projector version. This implementation allows for probing the correlation length of seeing perturbations across the telescope pupil, by varying the separation between the projectors. A common pulsed light source illuminates an array of collimated beams through a bifurcated fiber. Using N projectors allows for a tomographic probe of the beam deflection. If the beams can be arranged to intersect, the beam deflection from that parcel of air appears as common mode rather than differential motion in the focal plane. This could be a way to probe the dominant source of image motion along the boresight.}
    \label{fig:2cbp}
\end{figure}

We seek a configuration that can monitor image motion due to perturbations along the light path from the dome interior (or perhaps from just outside the enclosure slit) through the telescope and instrument optics and onto the main focal plane. Figures \ref{fig:rdimm} and \ref{fig:2cbp} show two implementations of the proposed concept that accomplishes this. Differential image motion is determined by looking at the variance in the distances between the centroids of the PSFs of the images of each collimated beam. This difference between pairs is a vector quantity, which allows for the computation of $\sigma_x^2$, $\sigma_y^2$, and $\sigma_x \sigma_y$. The differential-motion approach isolates the atmospheric effects from common-mode effects such as vibrations in the mount or mirrors.

Large format CCD cameras have shutters that are sluggish compared to the characteristic msec timescales of interest. While CMOS imagers cam deliver effective exposure times on this timescale, those are not yet the detector of choice for most astronomical instruments. Taking a multi-second long exposure suppresses the dynamic msec-timescale image motions introduced by local seeing, making it hard to distinguish that contribution to image FWHM from static figure errors and defocus. 

The solution we propose here is to use a fast-pulse light source to ``strobe'' the structure of the local atmosphere and obtain an effective short exposure time, while the main instrument's shutter is open. By projecting multiple diverging collimated beams into the main telescope, derived from a common pulsed source, variations in pulse intensity are not a source of uncertainty, nor are common-mode motions of the collimator or telescope. This method can clearly discriminate static sources of image degradation from local atmospheric effects. 

Operating with one flash per image on the main instrument is the simplest implementation, using a stack of images to determine image motion statistics. 

The advantage of this method over alternatives is the ability to probe directly the fluctuations in intensity and image centroid position due to optical path perturbations along the light path of the telescope. The short-pulse light allows us to take many realizations of the ``frozen'' atmosphere, even with an instrument shutter that is slow compared to the millisecond characteristic time scale. 

Figure \ref{fig:rdimm} shows a configuration that is effectively a strobed DIMM run in reverse. Optical wedges are placed across exit pupil sub-apertures from a collimated projector. The relative displacements of the centroids of the resulting images in the focal plane are a direct measurement of the image motion induced by local index perturbations. 

One amusing concept is to arrange, as shown conceptually in Figure \ref{fig:2cbp}, to have probe beams intersect at some position along the path. At that location the beams experience the same deflection, and the deflection from that parcel of air appears as common mode rather than differential motion in the focal plane. The amount of differential motion suppression as a function of beam intersection location could be a way to determine the location of the dominant seeing layer, by mapping out the differential to common mode motion ratio as a function of beam intercept location.  

\begin{figure}
    \centering
    \includegraphics[width=3.8in]{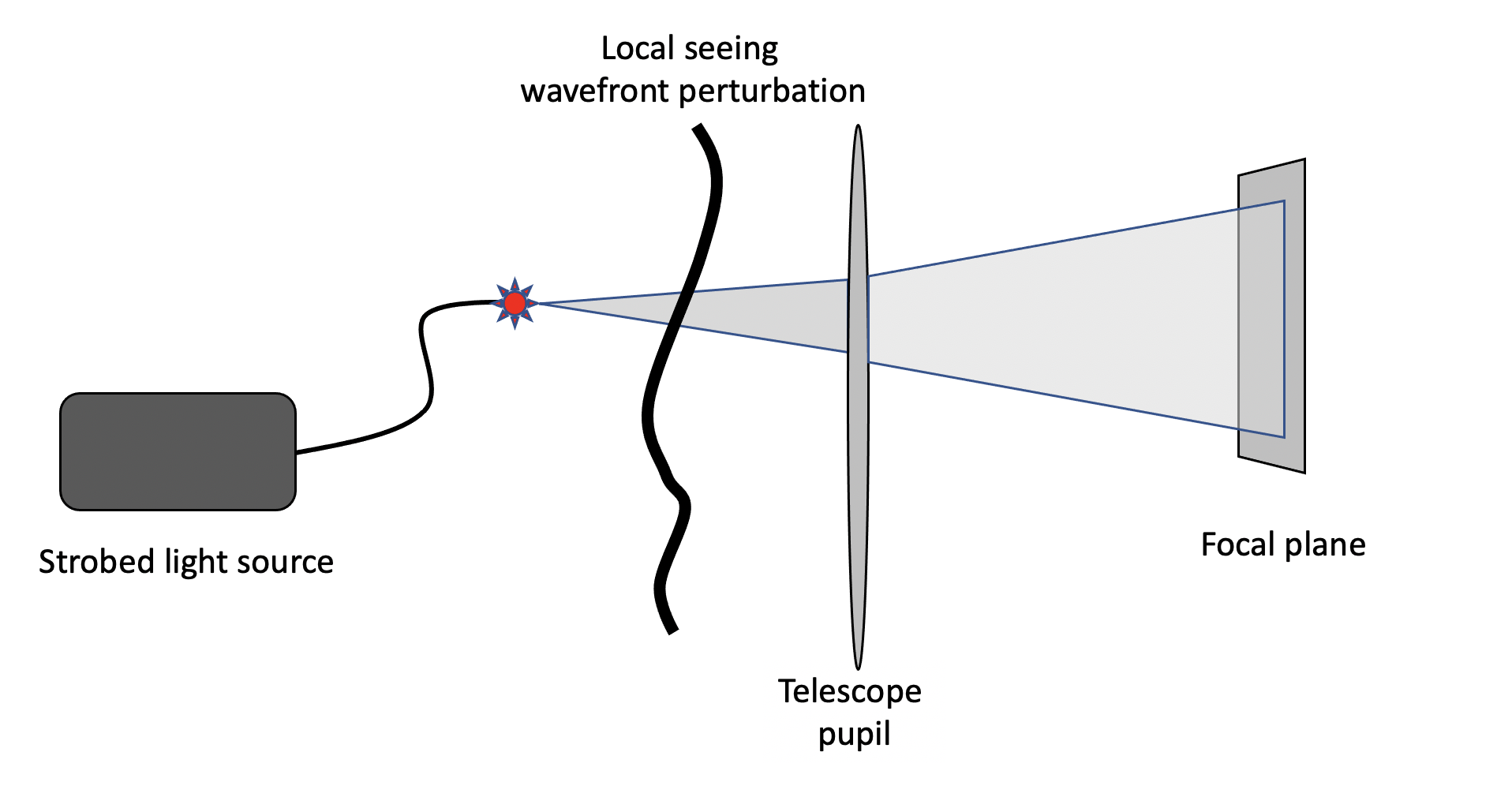}
\caption{Strobed Local Seeing Monitor, Scintillation Version. In order to probe scintillation along the light path inside the enclosure, a light from a low-divergence ($<$1 mrad) laser module with a small beam (few-mm diameter) is sent into the instrument. This beam size at the emitter is a good match to the Fresnel scale for scintillation on the focal plane. Pulsing the source for varying durations can allow for studies of temporal coherence, a useful diagnostic. A few-mW emitter with a beam divergence that matches the 3.5 degree Rubin focal plane should produce sufficient signal levels to allow for the determination of excess variance. }
    \label{fig:scintscan}
\end{figure}

Figure \ref{fig:scintscan} shows one conceptual design for the determination of scintillation along the line of sight within the dome. In this case we need a small emitter size, to avoid aperture-averaging effects. The illuminated footprint on the focal plane is determined by the divergence of the laser emitter. 

Scintillation conserves the energy in the beam that reaches the focal plane, it just re-arranges where the photons land. Pulse-to-pulse variation in the light source can be compensated by looking at the excess variance in the surface brightness of the illuminated area, after normalizing each image to a common integrated flux. 

\subsection{Engineering Parameter Choices}

In this section we explore some of the factors that influence engineering design choices. 

\subsubsection{Optical Wavelength and Passband}

The deflection of the rays is effectively achromatic, since the wavelength-dependence of the index of air is small. This has the advantage that we can operate the image-motion monitor in any wavelength or range of wavelengths that is advantageous. If the system performance is diffraction-limited there is merit to running at short wavelengths, with appropriate attention to UV safety issues.

For the scintillation monitor, there is merit to running sources at different wavelengths, to assess directly the wavelength-dependence of the scintillation along the path. In general the amplitude fluctuations are larger at shorter wavelengths. 

\subsubsection{Exposure Times, and Optical Pulse Duration}

The concept is to take a succession of images, each of which ``freezes'' the instantaneous structure of the optical index of refraction along the various propagation paths. Shorter pulses allow for sensitivity to higher temporal frequencies in the transverse direction. The characteristic seeing correlation time $\tau$ can be as short as msec. Exposures longer than this smear out the image motion, so we want a system capable of producing msec flashes. This is not technically challenging. The strobe duration is the effective exposure time, with a frame rate dictated by the shutter cycling and readout time of the system. We stress that the simultaneous-dual-sliding-curtain mode of obtaining short exposures on large cameras is inappropriate for this method. The shutter should be fully open before the strobe is triggered, to ensure all the transient light makes it to the focal plane. The ability to generate varying durations of illumination (ideally at fixed total photon dose) will allow us to measure the correlation time of the local seeing, by mapping out $\sigma^2$ vs. flash duration. It seems plausible that the power spectrum of image motion, probed by illuminating with pulses of varying duration, is different for mirror seeing vs. convection in the upper parts of the dome, for example. This might help in discriminating the dominant contribution to local seeing. The exploration of this speculation could involve driving exaggerated signatures of the various local seeing components, through intentionally large changes in ventilation and heat exchange. 

\subsubsection{Optical Pulse Energy and Peak Power}

If we require 100,000 detected photoelectrons per pulse per PSF in the focal plane to achieve good statistics, after accounting for mirror reflections and detector QE losses we need of order $10^6$ collimated emitted photons per pulse, independent of pulse duration. As shown below, a reasonable collimator aperture is 100mm, from which we desire a beam divergence of under an arcsecond. The (A-$\Omega)_{coll}$ product for the emitted beam is $\pi (0.1/2)^2 \times (5 \times 10^{-6})^2 = 2 \times 10^{-13} $ m$^2$ sr. 

If we use a high power LED source, with a 5mm x 5mm emitting into $\pi$ sr, the radiance is (A-$\Omega)_{LED} \sim$ 8 $\times 10^{-5} $ m$^2$ sr. The LED to collimated-beam efficiency is of order 10$^{-9}$. 

This implies we need to send a pulse of order $10^{15}$ photons down the fiber, to illuminate a few-micron pinhole at the collimator focus. For 500 nm light with $3 \times 10^{-19}$ Joules per photon  this corresponds to sending 0.3 mJ per pulse into the fiber. Packing all these photons into a 1 msec pulse duration, the shortest we're likely to need, requires a peak optical power of 0.3 W. 
 
Even feeding ten or more fibers, this power level should be readily achievable with a Xenon flash lamp or a pulsed LED. Driving LEDs in this regime is not challenging. A comparison of Xenon flash lamps vs. pulse-driven LED illuminators is provided by \cite{wilson2015performance}. 

For the scintillation monitor, the beam footprint spans a focal plane area that is determined by the source's beam divergence. A low-divergence laser diode typically has 0.25 mrad of divergence, for a diameter of just under an arcminute. This encompasses 2.83 $\times$ 10$^3$ square arc seconds, or 70 Kpixels on the Rubin focal plane. The energy impinging on the sensor for a 1 mW emitter is then roughly 1.5 $\mu$W per pixel, or a whopping 10$^{12}$ photons per second per pixel. If beam divergence is increased so that the beam spans the entire 3.5 degree x 3.5 degree focal plane, with 3.5 Gpixels, the flux per pixel becomes a more manageable 10$^6$ photons/pix/sec per mW emitted from the source.  A 1 msec strobed full-field image would produce around 1000 photons per pixel for a very modest 1 mW laser diode, which seems adequate to determine scintillation by looking for excess variance in flux levels.    

% \subsection{Optimizing Apertures}

\subsubsection{Collimator for Image Motion}
The uncertainty in centroid position is well represented by $\sigma_{centroid}=FWHM/SNR$, where the PSF is characterized by a width given by the Full Width at Half Maximum (FWHM) and SNR is the Poisson-limited signal to noise ratio for the image of the source. Small PSFs and high photon SNR are favorable for measuring centroid displacements. To address the regime of interest, we should have FWHM $<$ 1 arcsecond. This requires that the collimated projector produce a beam with a divergence below 5$\mu$rad. At the wavelengths of interest, the diffraction limit implies a collimator aperture $D$ of at least 100 mm. A diffraction-limited collimator produces a centroid uncertainty $\sigma_{centroid}$ that scales as D$^{-1}$, at fixed photon flux.

Trading against this is aperture-averaging of the beam deflection.  The RMS image motion (the signal of interest, see equation 1) scales as $D^{-1/6}$.

These scalings with collimator aperture indicate that the diffraction factor wins, with the SNR for centroid motion scaling as $D^{5/6}$, at fixed photon dose. 
This holds as long as larger collimator apertures produce smaller beam divergence, {\it i.e.}  while collimator diffraction determines the FWHM on the focal plane.     

The availability of high quality achromatic reflectors in the 8-12 inch aperture range suggest that size might be a good choice for the collimators. Larger optics become increasingly unwieldy and expensive.

\subsubsection{Scintillation Source}

The scintillation measurement favors a small aperture emitter, with $D<\sqrt{\lambda L}$ where $L$ is the optical path length from the source to the focal plane. We can adjust both the size and the divergence of the source to meet different goals. It's amusing that a source with a diameter at the pupil of $DP$ and divergence $\theta=DP/FL$, where FL is the focal length of the main telescope, will produce a spot on the focal plane of the same size, $DP$. This means the beam size stays essentially constant as it propagates through the focusing part of the optical system. That aside, mm beam sizes are readily available for low-divergence laser sources. As long as the speckle pattern arising from the source is temporally stable, frame subtraction of successive strobed images will suppress excess variance arising on the focal plane due to fixed laser speckle. 

% \subsubsection{Projector Support, Location, and Separation}

\section{Operational Concepts}

The simplest implementation of strobed-source wavefront monitoring would involve the following sequence: 
\begin{enumerate}
    \item Open instrument's main shutter,
    \item Enable either the image motion device or the scintillation source, or both. 
    \item Transmit optical pulses,
    \item Close instrument shutter,
    \item Read out the focal plane array, 
    \item Repeat multiple times, 
    \item Compute centroids and aperture photometry, and compute standard statistics on image motion and scintillation. The scintillation analysis will benefit from using frame subtraction to suppress any quasi-static speckle pattern from the coherent source. 
\end{enumerate}

More sophisticated implementations could include constant-readout or stuttered row shifts with the shutter open, imaging a succession of multiple illumination strobes. This would generate multiple PSF realizations in each image, allowing for high data collection efficiency. A cruder version of this would entail offsetting the telescope, with the shutter open, between multiple strobes. One could extend this concept to multiple pinholes in the collimator focal plane along with multiple wedged apertures in its output pupil, which 
would allow one to do tomography as described in \cite{hickson2019multistar,beltramo2019characterization}.

\subsection{Rubin Guiders and Wavefront Sensors as Local Seeing Monitors}
The four corners of the Rubin observatory focal plane each contain 4K x 4K ''guider'' CCDs, capable of reading out a sub-array at 10 Hz, as well as vertically offset wavefront sensors. Whether closed loop guiding will be implemented for the short (15-30 s) exposures anticipated for the Rubin system is an open question, to be determined during the commissioning phase. The four guiders lie 1.43 degrees off the telescope boresight. These four detectors could be used in conjunction with strobed sources to obtain high-data-rate statistics on image motion, using a strobed reverse-DIMM as shown in Figure \ref{fig:rdimm}. A pair of spatially separated beams could be projected into the pupil, using a collimator mounted to the top end of the telescope. An average off-axis angle of 1.43 degrees for the two beams, and the beams diverging from each other by an angle that spans the guider (a few arc-minutes) would provide a pair of in-focus centroids on the guider chip.  The collimated projector for this could be mounted on the top end of the telescope in an area obscured by the camera system and the secondary mirror. As long as the beams strike the primary mirror at the correct angle and are unobscured, they will land on the guider. 

Another option would be to flash-illuminate the wavefront sensors with one or more displaced beams, either instead of or in addition to the guider chips.  Since the wavefront sensors are only read out along with the science frame, small tweaks of beam directions between strobes could generate multiple realizations during a single science image. The Rubin wavefront sensors are each displaced by 2mm from the nominal focus of the Rubin telescope. The outer diameter of the full-pupil ``donut'' on each sensor therefore spans 1.67mm, and this corresponds to a distance of 8.5m at the input pupil. A cylindrical collimated beam of diameter DB entering the pupil will therefore span a distance  DWFS=1.67mm$~\times~$(DB/8.5m) = 20 $\mu m ~\times$ (DB/0.1m) = 2 pixels $\times$ (DB/0.1m) on the wavefront sensor. This should readily allow for the extraction of high SNR centroids from the resulting spots, even for a collimated illumination beam 100mm in diameter.

This illuminate-the-focal-plane-from-the-top-end configuration would allow for local seeing diagnostics to be interleaved with science observations, with no overhead for switchover. A more sporty version would operate the strobed source even during science integrations. While careful attention to stray and scattered light and ghosting would be important, the strobed collimated artificial source must have a brightness comparable that of the guide stars being contemplated, and should not produce any more scattered background light than those celestial objects. Tracking-rotation of the instrument can be compensated by rotating the collimator's exit pupil mask about the optical axis of the telescope, to continuously direct the beam(s) to one or more guider or wavefront CCDs.  

% \section{Proof of Principle}

\section{Discussion}

Strobed imaging for capturing transient events has a long history, from the legendary work of Edgerton and his contemporaries to modern studies of short-time-scale phenomena. Extensions of the idea presented here, namely the use of strobed imaging to capture local seeing, could include mast-mounted or drone-borne strobed sources, to extend the method to investigate boundary layer seeing outside the dome.  

The calibration plan for the Rubin observatory includes \citep{coughlin2016collimated} a collimated beam projector for monochromatic flux calibration. That device could be readily adapted to the scheme shown in Figure \ref{fig:rdimm}. 

The explicit measurement of image motion, with the correct weighting vs. location along the line of sight, can be used to explore the most advantageous ventilation configuration and thermal management in the dome, without having to discriminate upper atmosphere seeing or inferring image quality metrics from indirect measurements of turbulence or temperatures. Taken in conjunction with the determination of scintillation-induced variance as a function of flash duration and wavelength, these are potentially useful tools for characterizing the structure of the index of refraction within the enclosure, along the light path of primary interest. 

The airflow within a telescope enclosure strongly depends on the windspeed as well as azimuth difference between the wind direction and the enclosure slit. Wind-driven motion of the optical system depends on many factors, including windspeed and relative direction, and telescope elevation. An understanding of how these conditions influence local seeing could inform image-quality-optimized and condition-dependent scheduling of observations. 

At a minimum, with a projector mounted on the enclosure, the techniques described here should allow for daytime engineering tests of thermal and ventilation management, using a direct comparison of local seeing diagnostic data to system configuration and operational parameters. With a projector mounted on the telescope this could be extended to night-time tests with the slit open, perhaps even during science operations. 

Making the most of the substantial ongoing investments in ground-based telescopes is a high priority. Given how strongly the delivered image quality impacts scientific performance,
we need to identify and minimize local sources of image degradation. Strobed imaging provides one potential means for doing so.

\section*{Acknowledgments}

I am grateful to the US Department of Energy for their support of our optimization and precision calibration efforts for the Rubin Observatory, under DOE grant DE-SC0007881, and to Harvard University for support of our program. Chuck Claver provided, as always, insightful comments and suggestions. Paul Horowitz, E. Carrington Gregory, and John Tonry provided helpful comments on the manuscript.

%%%%%%%%%%%%%%%%%%%%%%%%%%%%%%%%%%%%%%%%%%%%%%%%%%
\section*{Data Availability}

No new data were generated or analysed in support of this research.

%%%%%%%%%%%%%%%%%%%% REFERENCES %%%%%%%%%%%%%%%%%%

% The best way to enter references is to use BibTeX:

\bibliographystyle{mnras}
\bibliography{DomeSeeingMNRAS.bib} % if your bibtex file is called example.bib

% Alternatively you could enter them by hand, like this:
% This method is tedious and prone to error if you have lots of references
%\begin{thebibliography}{99}
%\bibitem[\protect\citeauthoryear{Author}{2012}]{Author2012}
%Author A.~N., 2013, Journal of Improbable Astronomy, 1, 1
%\bibitem[\protect\citeauthoryear{Others}{2013}]{Others2013}
%Others S., 2012, Journal of Interesting Stuff, 17, 198
%\end{thebibliography}

%%%%%%%%%%%%%%%%%%%%%%%%%%%%%%%%%%%%%%%%%%%%%%%%%%

% %%%%%%%%%%%%%%%%% APPENDICES %%%%%%%%%%%%%%%%%%%%%

% \appendix

% \section{Some extra material}

% If you want to present additional material which would interrupt the flow of the main paper,
% it can be placed in an Appendix which appears after the list of references.

% %%%%%%%%%%%%%%%%%%%%%%%%%%%%%%%%%%%%%%%%%%%%%%%%%%

% Don't change these lines
\bsp	% typesetting comment
\label{lastpage}
\end{document}